 \documentclass[cameraready]{Interspeech}

\usepackage{comment}
\usepackage{tablefootnote,subcaption}
\usepackage{booktabs,multirow,makecell}

\graphicspath{{figures/}}

\title{Augmenting Dysarthric Speech Severity Assessment with MOS Supervision}

\makeatletter
\providecommand{\name}[1]{\gdef\authorlist{#1}}
\makeatother
\name{Kaimeng Jia$^\dagger$$^1$, Minzhu Tu$^{\dagger2}$, Zengrui Jin$^{*1}$, Siyin Wang$^1$, Chao Zhang$^1$ 
\thanks{$\dagger$ Equal contribution was made between the first two authors.}
\thanks{* Corresponding author.}
}

\address{
$^1$ Tsinghua University, $^2$ Beijing University of Posts and Telecommunications\\
\texttt{\small{jkm23@mails.tsinghua.edu.cn}}\small{,}
\texttt{\small{Epiphany\_1104@bupt.edu.cn}}\small{,} \\
\texttt{\small{wangsiyi23@mails.tsinghua.edu.cn}}\small{,}
\texttt{\small{\{zrjin,cz277\}@tsinghua.edu.cn}}
}

\email{}


\begin{document}
\maketitle
\begin{abstract}
Dysarthria is a speech disorder marked by reduced intelligibility and communicative effectiveness.
Automatic utterance-level assessment of dysarthric speech can support scalable speech monitoring and therapy-related analysis.
Yet training such systems is bottlenecked by the scarcity of clinically annotated dysarthric speech.
This work proposes to augment dysarthric speech assessment using data from speech synthesis evaluations, specifically human-annotated utterances with Mean Opinion Score (MOS) labels from the QualiSpeech corpus.
Experiments show that fine-tuning on speech synthesis assessment data consistently improves performance on both intelligibility and naturalness prediction, while joint training yields gains primarily on naturalness.
These results suggest that synthesis artifacts and dysarthric speech share perceptual commonalities, and speech synthesis evaluation corpora offer a practical augmentation source that reduces reliance on scarce clinical annotations.
\end{abstract}

\section{Introduction}
\label{sec:intro}

Dysarthria is a neuro-motor speech disorder resulting from neurological injuries or diseases,
such as cerebral palsy, amyotrophic lateral sclerosis, Parkinson's disease, or stroke
\cite{rudzicz2011tasl,rudzicz2009icassp}.
These disruptions manifest in several perceptually salient characteristics, with reduced intelligibility and degraded naturalness being among the most prominent and impactful on daily communication.


Despite recent advancement achieved in automatic speech recognition (ASR) \cite{hadian2018end, gulati20_interspeech, yao2024zipformer, yao2025crctc} and dysarthric speech recognition \cite{jin2023personalized, wang23y_interspeech, wang2024enhancing}, accurate assessment of dysarthric speech remains a challenging problem with important clinical and technological implications. 
It serves as a sensitive biomarker for early detection and tracking of neurological disease progression \cite{10095664}, provides an objective measure to guide speech therapy and monitor rehabilitation outcomes \cite{bhat2017icassp}, and is a key factor for improving downstream systems such as dysarthric speech recognition and disordered speech reconstruction \cite{geng23b_interspeech,jeon25_interspeech,10445949,wang2024unit}.

Despite its significant importance, clinical assessment of utterance-level intelligibility and
naturalness still relies heavily on subjective evaluations by certified speech pathologists
\cite{rudzicz2012torgo,kim08c_interspeech}, which are labor-intensive and difficult to scale.
These limitations have motivated the development of automated, objective assessment methods.
However, patients with speech disorders often present with co-occurring physical disabilities,
making large-scale data collection particularly challenging.
Furthermore, existing approaches are typically restricted to constrained lexicons and require
matched control groups, resulting in an unnatural evaluation protocol.
To the best of our knowledge, spontaneous utterances with unconstrained vocabulary have not
been employed in prior studies.

To address the data scarcity issue, self-supervised learning (SSL) has emerged as a powerful paradigm
in speech processing, offering robust and transferable representations from large-scale
unlabeled corpora. Models such as wav2vec 2.0 \cite{baevski2020wav2vec} and HuBERT
\cite{hsu2021hubert,yang2024k2ssl} have been successfully applied to both detection and
severity assessment of dysarthria \cite{yeo2023icassp,javanmardi2023wav2vec}.
A wide spectrum of data augmentation techniques has also been explored, including signal processing \cite{geng20_interspeech}, adversarial
domain adaptation \cite{10889800}, generative approaches
\cite{wang23qa_interspeech, jin21_interspeech,jin2023adversarial}, reverse autoencoders that transform healthy speech into
dysarthric speech \cite{bhat22_interspeech}, and other pre-trained TTS systems
\cite{hermann23_interspeech,leung24_interspeech,kim25w_interspeech}.
However, these generative approaches are designed to model spectro-temporal
characteristics of dysarthric speech for ASR or speech reconstruction purposes.
The augmented data they produce lacks perceptual validation, as no clinical experts
re-annotate the generated samples with severity labels, leaving it unclear whether
the synthetic speech aligns with human perceptual judgments.
As a result, such data cannot be directly used as augmentation for severity assessment,
where label quality and perceptual alignment are critical.

Meanwhile, automated assessment of Text-to-Speech (TTS) synthesis quality has been
extensively investigated \cite{huang22f_interspeech, lo19_interspeech, yang2025towards, yang2026towards, 11463437,wang2025enabling,chen2025audio,wang2025towards}, producing large quantities of
human-annotated utterances with mean opinion score (MOS) labels measuring naturalness and intelligibility.
Crucially, both synthesis artifacts and dysarthric speech represent deviations from
natural human production, sharing perceptual characteristics such as reduced
intelligibility, unnatural prosody, and degraded fluency.
This acoustic and perceptual overlap motivates a novel data augmentation strategy,
leveraging TTS assessment corpora, which are readily available and richly annotated,
as an augmentation source for dysarthric speech assessment systems.

The contributions of this work are threefold:
\textbf{(1)} An empirical study of SSL-based automatic dysarthric speech assessment on utterances with unconstrained vocabulary, going beyond the constrained protocols of prior work. 
\textbf{(2)} Demonstration that synthesized speech with human-annotated MOS scores from TTS evaluation corpora can effectively augment dysarthric speech assessment systems.
\textbf{(3)} Evidence that synthesis failures and dysarthric manifestations share perceptual and acoustic commonalities, offering a novel perspective on cross-domain knowledge transfer between TTS and articulatory disorders.

The rest of this paper is organized as follows.
Section~\ref{sec:task} introduces the assessment task and describes the Speech Accessibility Project (SAP) and QualiSpeech
corpora.
Section~\ref{sec:method} presents the model architecture and the two training paradigms:
joint training and fine-tuning.
Section~\ref{sec:exp} reports experiments and results on intelligibility and naturalness
prediction.
The last section concludes this work.

\section{Task Description}
\label{sec:task}

\subsection{The Speech Accessibility Project Challenge 2025}

\begin{figure}[ht]
  \centering
  \begin{subfigure}[b]{0.49\columnwidth}
    \centering
    \includegraphics[width=\columnwidth]{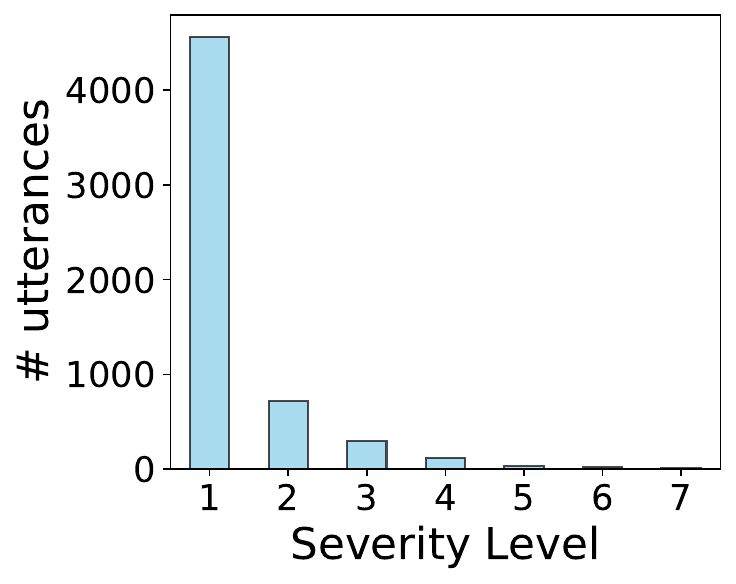}
    \caption{Intelligibility}
    \label{fig:intelligibility_dist}
  \end{subfigure}
  \hfill
  \begin{subfigure}[b]{0.49\columnwidth}
    \centering
    \includegraphics[width=\columnwidth]{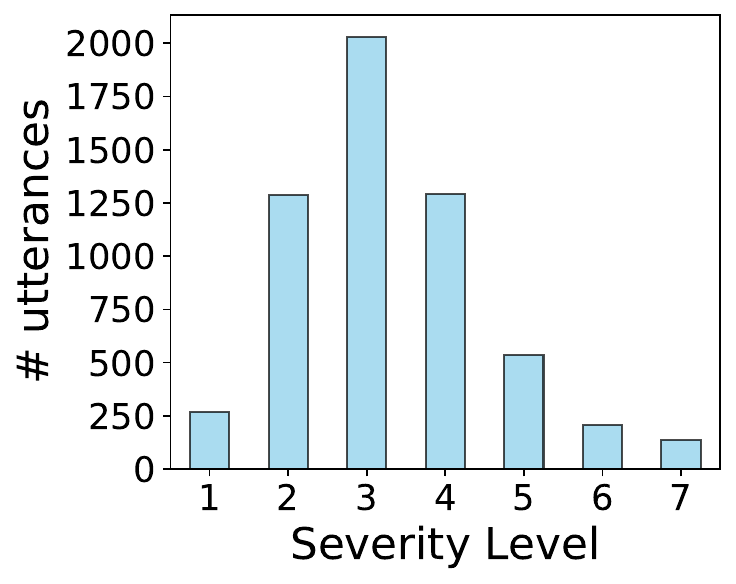}
    \caption{Naturalness}
    \label{fig:naturalness_dist}
  \end{subfigure}
  \caption{Distribution of utterances across severity levels for \textbf{(a)} Intelligibility and \textbf{(b)} Naturalness in the Speech Accessibility Project (SAP) challenge. Intelligibility is heavily biased toward level 1 (minimal impairment), while Naturalness exhibits a more balanced spread across levels, reflecting the greater variability in perceived speech quality among dysarthric speakers.}
  \label{fig:distribution}
\end{figure}

The SAP challenge \cite{zheng25_interspeech} provides a large-scale, open-domain corpus of dysarthric speech.
The corpus comprises utterances recorded from over 500 speakers diagnosed with Parkinson's disease, Down syndrome, amyotrophic lateral sclerosis, Cerebral palsy, or stroke, encompassing more than 400 hours of speech and over 190,000 utterances.
The dataset was recorded through participants using their personal devices at home, capturing both read and spontaneously generated speech with unconstrained vocabulary to ensure naturalness and variability.
Fine-grained perceptual speech assessment across multiple clinically relevant dimensions is provided in the corpus, including but not limited to, harsh voice, inappropriate silences, pitch level, naturalness, variable rate, and intelligibility. 
Each dimension is annotated by certified speech-language pathologists using a 7-point scale, where a higher score indicates more significant severity is manifested in the corresponding dimension. 
Given the complexity among these perceptual dimensions and the primary objective of dysarthric speech assessment, the focus is specifically narrowed to two core utterance-level annotations, Intelligibility and Naturalness.
Distribution of utterances across severity levels can be seen from Figure \ref{fig:distribution}.

Due to the unavailability of the original SAP test labels, the development partition was used as the test set. 
The released SAP training and development partitions utilize a speaker-level data partitioning protocol, thus the test speakers are unseen during SAP training. 
For each target dimension, utterances annotated for that dimension were selected from the SAP training partition, yielding 5,046 utterances for Intelligibility and 5,040 for Naturalness before validation sampling. 
Validation sets consisting of 500 utterances per dimension were randomly sampled from the corresponding training subsets and used only for model selection. 
These validation utterances were removed from the corresponding training subsets, ensuring no utterance overlap across train, validation, and test. 
The resulting test sets include 716 utterances for Intelligibility and 714 utterances for Naturalness.

\begin{figure*}[ht]
  \centering
  \includegraphics[width=\textwidth]{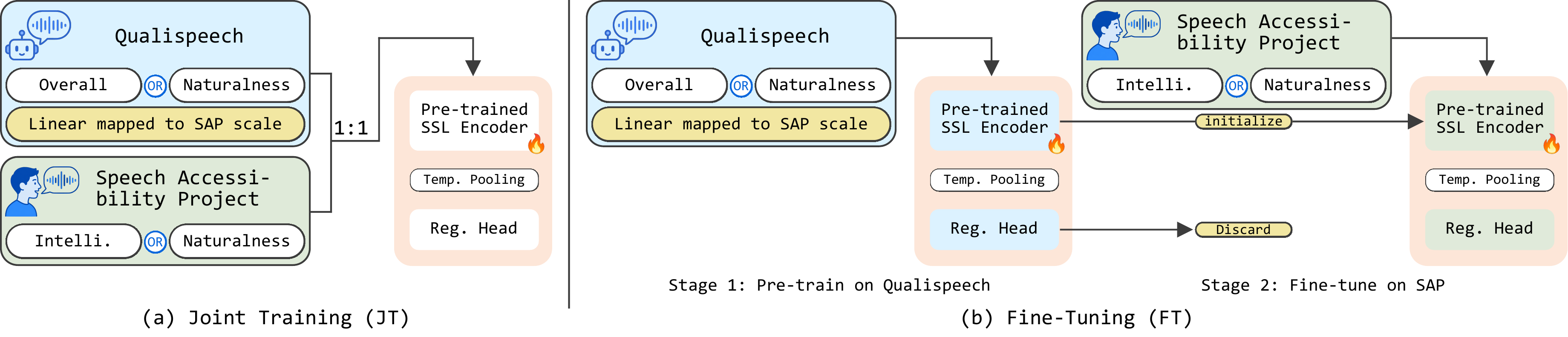}
  \caption{An illustration of the proposed training paradigms. In Joint Training (JT), QualiSpeech and Speech Accessibility Project (SAP) utterances are mixed at a 1:1 ratio and fed jointly into a shared SSL encoder and regression head, with QualiSpeech MOS scores linearly mapped to the SAP severity scale. In Fine-Tuning (FT), the model is first trained on QualiSpeech, then fine-tuned on SAP without simultaneous cross-dataset optimization.}
  \label{fig:framework}
\end{figure*}

\subsection{QualiSpeech: A Descriptive Corpus for Speech Quality Assessment}
QualiSpeech \cite{wang-etal-2025-qualispeech} is an English corpus for low-level speech quality assessment with comprehensive annotations and detailed descriptive comments. 
It comprises three speech categories: synthetic speech (49\%), primarily from the Blizzard Challenge and Voice Conversion Challenge (BVCC) corpus \cite{cooper21_ssw} and various TTS models with standardized MOS scores; simulated real speech (27\%), sourced from the Non-Intrusive Speech Quality Assessment (NISQA) corpus \cite{mittag21_interspeech} featuring distortions that emulate actual transmission; and real speech (24\%), encompassing NISQA live recordings and in-the-wild samples from GigaSpeech \cite{chen21o_interspeech}. 
To ensure balanced speech quality conditions, 20\% of the synthetic speech samples are mixed with noise at signal-to-noise ratios ranging between 0 and 15 dB, while real speech samples are stratified into four quality groups based on UTMOS-predicted MOS scores \cite{saeki22c_interspeech}. 
Each utterance is evaluated across seven perceptual quality dimensions, namely noise, distortion, speed, continuity, listening effort, naturalness, and overall quality.
This protocol results in training, validation, and test splits of 10,558, 2,167, and 1,852 utterances, respectively, with balanced compositions of synthetic and real speech. 

Among the seven annotated dimensions, overall quality and naturalness are selected as they provide complementary yet comprehensive perspectives on perceptual evaluation. 
Overall quality captures the aggregated listener impression across multiple degradations, while naturalness reflects the degree to which speech resembles genuine human production. 
Focusing on these two dimensions ensures both clinical relevance for pathological speech assessment and consistency with established practices in general speech quality evaluation.

\section{Method}
\label{sec:method}

\subsection{Model Architecture}

To leverage transferable representations from large-scale unlabeled speech, self-supervised learning (SSL) pre-trained encoders are adopted as the backbone for feature extraction \cite{cooper2022generalization}.
The core idea is to augment the limited in-domain dysarthric speech data with human-annotated TTS assessment data from QualiSpeech, using either joint training or fine-tuning to transfer perceptual supervision into the dysarthria assessment model.
Given a raw waveform input, the SSL encoder produces a sequence of frame-level contextual representations. 
These are aggregated via mean pooling over the time dimension to obtain a fixed-dimensional utterance-level embedding, which captures global perceptual characteristics without requiring explicit segmentation.
The utterance embedding is then passed to a regression head consisting of a two-layer feed-forward network with ReLU activation and dropout regularization between layers, producing a single continuous severity score.
The entire model, including the SSL encoder, is fine-tuned in an end-to-end fashion during training.

\subsection{Training Paradigms}

Two training paradigms are proposed to incorporate perceptual supervision from QualiSpeech into dysarthric speech assessment, as illustrated in Figure~\ref{fig:framework}.

\textbf{Joint Training (JT).}
The JT paradigm trains a single model simultaneously on both the QualiSpeech and SAP corpora under a unified regression objective.
To prevent the larger QualiSpeech corpus from dominating training, utterances are randomly sampled from the QualiSpeech training set to match the size of the SAP training split, yielding a balanced 1:1 mixture ratio.
Since the two datasets adopt different rating scales, with QualiSpeech using a MOS scale of 1--5 while SAP uses a severity scale of 1--7, a linear transformation is applied to align QualiSpeech scores before joint optimization:
\begin{equation}
    \hat{s} = 1 + (5 - s_{\text{MOS}}) \cdot \frac{6}{4},
    \label{eq:scale_align}
\end{equation}
where $s_{\text{MOS}} \in [1, 5]$ is the original QualiSpeech MOS and $\hat{s}$ is the aligned score in the SAP scale. 
Under this mapping, a MOS of 5 (highest quality) corresponds to an SAP score of 1 (no dysarthric characteristics), and a MOS of 1 (lowest quality) maps to an SAP score of 7 (most severe).

\textbf{Fine-Tuning (FT).}
The FT paradigm decouples learning into two sequential stages.
The model is first trained on QualiSpeech to predict the selected MOS dimension and acquire perceptual quality representations.
The resulting weights then initialize the SAP model, which is fine-tuned for dysarthric speech severity prediction.
This design transfers perceptual supervision while avoiding simultaneous cross-dataset optimization.

\subsection{Evaluation Metrics}

Model performance is evaluated using three complementary metrics: Mean Squared Error (MSE), Linear Correlation Coefficient (LCC), and Spearman's Rank Correlation Coefficient (SRCC).
MSE quantifies the absolute deviation between predicted scores and ground-truth ratings, with lower values indicating higher prediction accuracy.
LCC measures the strength of the linear relationship between predictions and references, while SRCC assesses the consistency of relative ranking order, which is particularly relevant for severity-level discrimination.
Together, these three metrics provide a comprehensive view of both regression accuracy and perceptual alignment with human judgments.

\section{Experiments and Results}
\label{sec:exp}

To leverage human-perception-aligned supervision from QualiSpeech for dysarthria assessment, we evaluate two TTS-based data augmentation paradigms: joint training (JT) and fine-tuning (FT).
All models evaluated are optimized using the Adam optimizer with mean squared error (MSE) as the training objective, a learning rate of $1e-5$, and weight decay of $0.01$. 
Details of the utilized self-supervised learning (SSL) pre-trained encoders are listed in Table \ref{tab:fairseq_models}.

\begin{table}
  \centering
  \caption{Details of self-supervised learning (SSL) pre-trained encoders.}
  \label{tab:fairseq_models}
  \resizebox{\linewidth}{!}{
  \begin{tabular}{l c l}
    \toprule
    Model &  \# Param. & Dataset   \\
    \midrule
    wav2vec 2.0 Base\tablefootnote{\url{https://dl.fbaipublicfiles.com/fairseq/wav2vec/wav2vec_small.pt}} & 94M  & Librispeech \cite{panayotov2015librispeech}      \\
    wav2vec 2.0 Large*\tablefootnote{\url{https://dl.fbaipublicfiles.com/fairseq/wav2vec/wav2vec_vox_new.pt}} & 315M & Libri-Light \cite{kahn2020libri}  \\
    wav2vec 2.0 Large+\tablefootnote{\url{https://dl.fbaipublicfiles.com/fairseq/wav2vec/w2v_large_lv_fsh_swbd_cv.pt}} & 315M & \makecell[l]{Libri-Light \cite{kahn2020libri} + CommonVoice \cite{ardila-etal-2020-common} + \\ Switchboard \cite{godfrey1992switchboard} + Fisher \cite{cieri-etal-2004-fisher}} \\
    \midrule
    HuBERT Base\tablefootnote{\url{https://dl.fbaipublicfiles.com/hubert/hubert_base_ls960.pt}} & 95M  & Librispeech \cite{panayotov2015librispeech}  \\
    HuBERT Large\tablefootnote{\url{https://dl.fbaipublicfiles.com/hubert/hubert_large_ll60k.pt}} & 316M    & Libri-Light \cite{kahn2020libri} \\
    \bottomrule
  \end{tabular}
  }
\end{table}

All training paradigms share an identical experimental setup with parameters of SSL encoders kept trainable.
Under the JT paradigm, 4,000 utterances are randomly sampled from the QualiSpeech training set with the original data distribution preserved, and then combined with the SAP training set for joint optimization. 

\begin{table*}[ht]
  \centering
  \setlength{\tabcolsep}{3pt}

  \caption{Results of self-supervised learning (SSL) pre-trained encoders under joint training (JT) and fine-tuning (FT) on the Speech Accessibility Project dysarthric speech corpus (SAP) and QualiSpeech. The two ``Dimension'' columns denote the SAP target dimension (left) and the QualiSpeech auxiliary supervision dimension (right). ``IDT'' denotes in-domain training on SAP only, with no QualiSpeech augmentation (``--''). For each SSL encoder, Mean Squared Error (MSE $\downarrow$), Linear Correlation Coefficient (LCC $\uparrow$), and Spearman's Rank Correlation Coefficient (SRCC $\uparrow$) are reported. \textbf{Bold values} indicate the best result per encoder within each SAP dimension group.}
  \label{tab:qs_sap_results}

  \resizebox{\linewidth}{!}{
  \begin{tabular}{c c cc *{5}{ccc}}
    \toprule
    \multirow{2}{*}{\textbf{ID}} &
    \multirow{2}{*}{\textbf{Method}} &
    \multicolumn{2}{c}{\textbf{Dimension}} &
    \multicolumn{3}{c}{wav2vec 2.0 Base} &
    \multicolumn{3}{c}{wav2vec 2.0 Large*} &
    \multicolumn{3}{c}{wav2vec 2.0 Large+} &
    \multicolumn{3}{c}{HuBERT Base} &
    \multicolumn{3}{c}{HuBERT Large} \\
    \cmidrule(lr){3-4} \cmidrule(lr){5-7} \cmidrule(lr){8-10} \cmidrule(lr){11-13} \cmidrule(lr){14-16} \cmidrule(lr){17-19}
       & & SAP & QualiSpeech
       & MSE & LCC & SRCC
       & MSE & LCC & SRCC
       & MSE & LCC & SRCC
       & MSE & LCC & SRCC
       & MSE & LCC & SRCC \\
    \midrule
    \midrule
    1 & IDT & Intelligibility & --
        & 0.348 & 0.628 & \textbf{0.482}
        & 0.421 & 0.523 & 0.368
        & 0.547 & 0.471 & 0.322
        & 0.461 & 0.540 & \textbf{0.406}
        & 0.475 & 0.485 & 0.404 \\
    \midrule
    2 & FT & \multirow{2}{*}{Intelligibility} & \multirow{2}{*}{Overall}
        & \textbf{0.272} & \textbf{0.751} & 0.427
        & \textbf{0.351} & \textbf{0.612} & 0.393
        & \textbf{0.308} & 0.534 & 0.285
        & 0.487 & \textbf{0.594} & 0.368
        & 0.431 & 0.596 & 0.356 \\
    3 & JT & &
        & 0.408 & 0.590 & 0.446
        & 0.560 & 0.317 & 0.330
        & 0.627 & 0.225 & 0.302
        & 0.612 & 0.387 & 0.317
        & 0.526 & 0.479 & 0.363 \\
    \midrule
    4 & FT & \multirow{2}{*}{Intelligibility} & \multirow{2}{*}{Naturalness}
        & 0.303 & 0.648 & 0.464
        & 0.387 & 0.572 & \textbf{0.401}
        & 0.495 & \textbf{0.566} & \textbf{0.443}
        & \textbf{0.379} & 0.451 & 0.295
        & \textbf{0.379} & \textbf{0.608} & \textbf{0.464} \\
    5 & JT & &
        & 0.433 & 0.602 & 0.475
        & 0.604 & 0.383 & 0.352
        & 0.528 & 0.445 & 0.348
        & 0.489 & 0.515 & 0.309
        & 0.451 & 0.551 & 0.391 \\

    \midrule
    \midrule
    
    6 & IDT & Naturalness & --
        & 1.127 & 0.581 & 0.591
        & 1.354 & 0.469 & 0.481
        & 1.119 & 0.574 & 0.607
        & 1.169 & 0.546 & 0.521
        & 0.941 & 0.570 & 0.503 \\

	\midrule

    7 & FT & \multirow{2}{*}{Naturalness} & \multirow{2}{*}{Overall}
        & 1.053 & \textbf{0.718} & \textbf{0.723}
        & 1.033 & 0.695 & 0.706
        & 1.000 & 0.686 & \textbf{0.696}
        & \textbf{0.847} & 0.644 & 0.587
        & \textbf{0.819} & \textbf{0.701} & \textbf{0.690} \\
    8 & JT & &
        & 0.909 & 0.691 & 0.680
        & 1.027 & 0.606 & 0.613
        & 1.106 & 0.589 & 0.617
        & 1.050 & 0.614 & 0.634
        & 0.930 & 0.661 & 0.668 \\
    \midrule
    9 & FT & \multirow{2}{*}{Naturalness} & \multirow{2}{*}{Naturalness}
        & \textbf{0.717} & 0.717 & 0.657
        & \textbf{0.800} & \textbf{0.709} & \textbf{0.713}
        & \textbf{0.880} & \textbf{0.692} & 0.690
        & 1.075 & \textbf{0.646} & \textbf{0.635}
        & 0.823 & 0.678 & 0.675 \\
    10 & JT & &
        & 1.022 & 0.637 & 0.643
        & 1.085 & 0.629 & 0.648
        & 1.022 & 0.637 & 0.642
        & 1.121 & 0.586 & 0.590
        & 1.036 & 0.638 & 0.635 \\
    \bottomrule
  \end{tabular}
  }
\end{table*}

\subsection{Performance on the Intelligibility Dimension}

Table \ref{tab:qs_sap_results} (Row 1–5) reports intelligibility prediction on SAP, with Overall quality and Naturalness from QualiSpeech used as auxiliary supervision under fine-tuning (FT) and joint training (JT) across five SSL encoders.
Three key trends emerge:

\noindent {\bf (i)} Under in-domain training (IDT, Row 1), wav2vec 2.0 Base achieves the strongest performance, with a 36.4\% relative MSE reduction over Large+, suggesting that smaller encoders generalize better to this task without additional supervision.

\noindent {\bf (ii)} Speech synthesis augmentation via FT (Row 2 vs. 1) consistently improves intelligibility prediction across all wav2vec encoders, with the largest relative MSE reduction of 43.7\% for Large+ and a 19.6\% LCC improvement for Base, demonstrating that QualiSpeech overall quality provides effective perceptual transfer for intelligibility.

\noindent {\bf (iii)} FT consistently outperforms JT across all encoder and dimension combinations (Row 2 vs. 3, Row 4 vs. 5), with Overall quality yielding the most robust MSE and LCC gains while Naturalness further improves SRCC, particularly for larger encoders, indicating that the choice of augmentation dimension interacts with model capacity.

\subsection{Performance on Naturalness Dimension}
Table \ref{tab:qs_sap_results} (Row 6–10) reports naturalness prediction on SAP, utilizing Overall quality and Naturalness from QualiSpeech as auxiliary supervision under fine-tuning (FT) and joint training (JT) across five SSL encoders.
Three key trends emerge:

\noindent {\bf (i)} Speech synthesis augmentation via FT (Row 7 vs. 6) consistently improves naturalness prediction across all encoders, with wav2vec 2.0 Large* showing the largest LCC (48.2\% relative) and SRCC (46.8\% relative) gains, confirming that overall quality from QualiSpeech transfers effectively to dysarthric naturalness assessment.

\noindent {\bf (ii)} Unlike intelligibility, JT also yields consistent gains over IDT on naturalness (Row 8 vs. 6), with wav2vec 2.0 Base and Large* reducing MSE by 19.4\% and 24.2\% respectively, suggesting that the perceptual alignment between QualiSpeech quality ratings and dysarthric naturalness is sufficient to support joint optimization.

\noindent {\bf (iii)} Using QualiSpeech Naturalness as auxiliary supervision under FT (Row 9) achieves the lowest MSE across all conditions, with wav2vec 2.0 Base and Large* achieving 36.4\% and 40.9\% relative MSE reductions, indicating that dimension-matched supervision yields the strongest transfer.

Taken together, these results demonstrate that the perceptual overlap between speech synthesis quality and dysarthric naturalness is sufficient to support effective cross-domain augmentation under both FT and JT, validating the use of speech synthesis evaluation corpora as a practical and label-efficient augmentation source for dysarthric speech assessment.

\subsection{Analysis on the Performance Gap between FT and JT}
The consistent gap between FT and JT on intelligibility can be attributed to cross-domain label misalignment and negative transfer under joint optimization.
The QualiSpeech supervision signals, overall quality and naturalness, are semantically misaligned with SAP intelligibility: while the latter reflects clinical judgments of phonemic clarity and word-level recoverability, the former capture aggregated listener impressions of multi-dimensional degradations and human-likeness, which are more closely related to the naturalness perceptual axis than to articulatory intelligibility.

Under JT, a shared regression head and a unified MSE objective are applied over both corpora, implicitly treating the linearly mapped QualiSpeech scores as a proxy for SAP intelligibility severity.
This assumption fails for intelligibility, as gradients from QualiSpeech bias the shared representations toward explaining MOS variance, suppressing the discriminative acoustic cues required for intelligibility assessment.

FT sidesteps this conflict by decoupling the two learning stages: QualiSpeech pre-training initializes the encoder with perceptually informed representations, while subsequent fine-tuning on SAP allows the model to re-weight features toward the target domain without cross-dataset gradient interference, explaining its consistent advantage over JT.
These findings suggest that speech synthesis evaluation data can serve as a viable augmentation source for dysarthric intelligibility assessment, provided that the transfer is mediated through sequential fine-tuning rather than joint optimization.

\section{Conclusion}
This study demonstrated that leveraging perceptual annotations from the QualiSpeech corpus significantly enhances automatic dysarthria assessment. 
Fine-tuning consistently improved prediction accuracy over the baseline on both dimensions, while joint training yielded gains primarily on naturalness.
Larger models derived greater benefit from additional supervision, and naturalness prediction consistently achieved stronger correlation with human judgments than intelligibility, likely attributable to the severe class imbalance in the intelligibility dimension as shown in Figure~\ref{fig:distribution}.
The effectiveness of synthetic speech augmentation underscores perceptual and acoustic commonalities between synthesis failures and dysarthric speech, highlighting promising directions for future clinical research.

\section{Generative AI Use Disclosure}
During the preparation of this manuscript, the authors used generative AI to correct grammar mistakes and misspellings.
Illustrative icons in Figure \ref{fig:framework} are AI generated. 
After using these tools, the authors carefully reviewed and edited the manuscript, and take full responsibility for the final content of the paper.

\bibliographystyle{IEEEtran}
\bibliography{refs.bib}

\end{document}